\documentclass[fleqn]{article}  % twocolumn % affiliation des auteurs
\usepackage[utf8]{inputenc}
\usepackage[T1]{fontenc}
\usepackage{graphicx}
\usepackage{amsmath}
\usepackage{amssymb}
\usepackage{amsfonts}
\usepackage{esvect}
\usepackage{ifthen}
\usepackage{mathtools}
\usepackage{kpfonts}
\title{Earth-skimming Ultra-high Energy Tau Neutrinos simulated with MonteCarlo method and CONEX code}
\author{Bouzid BOUSSAHA*, Tarek BITAM**}
\date{}
\begin{document}
\maketitle{}
%\begin{multicols}{2}
*University of Algiers, Algeria\\
**Université Djilali Bounaama Khemis Miliana,Algeria
\begin{abstract}
This paper aims to study the feasibility of building an Earth-skimming cosmic tau neutrinos detector, with the aim of eventually identifying the ideal dimensions of a natural site mountain-valley for the detection of very high energy neutrinos tau range from $ 10^{16} eV $ to $ 10^{20} eV $, as well as possibly locate one such site in Algeria. First, a Monte Carlo simulation of the neutrino-[mountain]matter interaction as well as the resulting decay of the tau lepton is conducted to determine the optimal dimensions of the mountain as well as the location of the tau decay in the valley. Second, a CORSIKA (COsmic Ray Simulation for KAscade) [1] simulation with the CONEX option is conducted to track the evolution of the almost horizontal air shower initiated by the tau lepton. Many particles are produced, wich are part of the shower components : electrons, muons, gammas, pions, etc. The study of the spatial distribution of these particles enables the discovery of the optimal width of the valley, and consequently, the distance at which to lay the detection network [15, 16].
\end{abstract}
\section*{Introduction}
Ultra High Energy (UHE) tau neutrinos are produced by neutrino oscillation phenomena of other flavours (electron and muon neutrinos) during of their journey from the creation site in the cosmos to the arrival point on Earth. The cross section for the neutrino interaction with matter, in this case the rock of Earth, increases with energy [2], i.e., the prospect of an interaction with matter is higher in the case of a UHE neutrino compared to solar neutrinos.
Moreover, Earth matter crossed by neutrinos does not meet the required size for the detection of low and medium energy neutrinos, which is advantageous in our case. The study of Earth-skimming UHE tau neutrinos was proposed as a method to apply in the search for neutrinos of cosmic origin by several research studies around the year 2000 [3, 4, 5, 6]. 
Various promising experiments for the detection of neutrino tau at higher energies through tau-induced air showers are developing and using Cherenkov telescopes, we cite: GRAND [7] , BEACON [8] and TRINITY [9], who shows that a small system of detection is sufficient to reach sensitivities that comparable with ARA [10] or ARIANA [11]experiments. Future experiments aims to detect up-going air showers observed from orbital and sub-orbital altitudes POEMMA [12], EUSO-SPB2 [13].\\
The device idea is to exploit mountain matter in order to convert (weak interaction) tau neutrinos into charged tau leptons, part of which emerges to decay in the atmosphere following an approximately horizontal direction, and forming an air shower. Some of the particles in the shower are detectable using an array of scintillation detectors coupled with photomultipliers, some of which have already been used for the detection of cosmic muons [14]. 
Subsequent to a judicious choice of the mountain, positioning the detection site close to the maximum development of the air shower will serve to optimize the detection. It should be noted that this configuration has already been proposed by the LPSC team at Grenoble [15, 16].

\section{Description of the experiment}
Consider a tau neutrinos $ \nu_{\tau} $ propagating at a given angle and arriving at the front of a mountain. It is then likely to interact with the mountain by weak interaction and, in some cases, create a lepton $ \tau $, which will then in turn propagate. If the lepton $ \tau $ is created at a sufficiently close distance to the surface of the mountain, it will have a high probability of escape and, through decay, trigger a horizontal air shower in the valley of the mountain.
This shower will then be detectable under the condition that the decay does not happen at too great a distance. To this purpose, a detection network is installed on a second mountain with the prerequisite that it is to be positioned to coincide with the maximum development of the air shower. The tau neutrino detector offers such detection through an installation in a specific mountain-valley configuration.
Similar to the tau neutrinos $ (\nu_{\tau}) $, a cosmic electron neutrinos $( \nu_{e}) $ is able to propagate in rocks.
However, the electron which is created via charged current interaction $( \nu_{e})N $, is too quickly stopped for it to be able to contribute significantly in this type of detection. On the other hand, a muon created during a $( \nu_{\mu})N $ interaction is able to escape from the rock, as is the case with $\tau$. However, the difference in this instance lies within the lifetimes of these two particles. Indeed, if the $ (\nu_{\tau}) $ has an extremely short lifetime of $2.9.10^{-13} s $, the muon has a sufficiently longer lifetime of 2.2 $\mu s$. Thus, at a given energy, the probability of muon decay in the valley is negligible compared to that of $\tau$. Similarly  to the $\nu_{e}$, the $( \nu_{\mu})$ will be undetectable by the device subject to this research.
To describe such a phenomenon in a precise manner, the following processes have to be considered :

$$
\textit{The neutrino}\left\lbrace
	\begin{aligned}
\textit{Weak charged current CC interaction}  \\
\textit{Weak neutral current NC interaction} 
	\end{aligned}
	\right.
$$

$$
\textit{The $\tau$ lepton}\left\lbrace
	\begin{aligned}
&\textit{Weak CC and NC interactions}  \\
&\textit{Energy loss: Ionization, radiative processes} \\
&\textit{Decay} 
	\end{aligned}
	\right.
$$

\section{Aims of study}
The objective of this study is twofold. First, identify a mountain of ideal width which will enable the production of a maximum number of emerging tau  leptons from the incident tau neutrinos.
Second, define the ideal width of the valley of the tau neutrinos conversion mountain. This width is to be determined by allowing the decay of the tau lepton and the subsequent detection of the resulting air shower at its depth of maximum (longitudinal development of the air shower). It will then be necessary to compare these sites dimensions to the existing natural sites in the country: mountains of a width between ten and a few tens of kms. In this study, mountain widths of up to an upper limit of 30 km and valley widths of up to 20 km were considered.

\section{Monte-Carlo description of the propagation inside the mountain and
valley}
\subsection{Inside the mountains: the conversion from Tau neutrinos to Tau leptons}
The interactions to be considered for the purposes of this study and the research question are : weak neutrino interactions; the processes of tau energy loss and decay. Given the energies involved in the study of cosmic tau neutrinos (above $ 10^{16} eV $), it is possible to regard the direction of both the emerging and the incident tau leptons as approximately the same, and thus, a single-dimensional study should be sufficient.
The interaction position of a neutrino is drawn randomly when it is created, which is much faster than the step-by-step propagation. With regard to the neutrino which has only two possible interactions and which does not experience any energy loss, the distance between its point of creation and its point of interaction is given by [17].

\begin{equation}
$$
D_{int}=\frac{-\ln(R)}{N_{A}\times\rho\times(\sigma_{CC}(E)+\sigma_{CN}(E))}
$$
\end{equation}

Where R is a random number between 0 and 1; $\sigma_{CC}$ and $\sigma_{CN}$ are the cross sections of charged current and neutral current interactions, respectively, given by formula(2)[18,19].

\begin{equation}
$$
\sigma_{CC}(E)=2.4\times\sigma_{CN}(E)=6.04\times10^{-36}cm^{2}(\frac{E}{GeV})^{0.358}
$$
\end{equation}

In the case where a charged current interaction occurs (ex : $\nu_{\tau} n \rightarrow \tau^{-} p$), i.e., an interaction producing a tau lepton, we will follow the propagation of the resulting $\tau$ instead, by determining
its energy loss at each step, as well as the probability of its decay. In the event of decay, the resulting neutrinos will then propagate again. It may eventually produce tau (A process called "neutrinos regeneration"). These events are denominated "double-bang" events (and singleconversion event "single-bang" events). The probability of "double-bang" events is however
negligible [17].\\
The interaction position of the neutrino can therefore be randomly derived. If the position is outside the rock, the neutrino is considered as lost. If there is an interaction in the Earth, then its type (of the two possible interactions) has to be determined.
Due to the energy losses to be applied, tau lepton should normally be propagated step by step. Alternatively, the decay position of the tau lepton, is also randomly derived according to the following formula for the survival probability of the particle given by Dutta, Huang and
Reno [20]:

\begin{equation}
$$
P_{surv}(E_{\tau},E_{\tau}^{i})=\exp\Bigg[\frac{m_{\tau}\beta_{1}}{c\tau\rho\beta_{0}^{2}}\bigg(\frac{1}{E_{\tau}}(1+\ln(\frac{E_{\tau}}{E_{0}}))-(\frac{1}{E_{\tau}^{i}}(1+\ln(\frac{E_{\tau}^{i}}{E_{0}}))\bigg)\Bigg]\exp\Bigg[-\frac{m_{\tau}}{c\tau\rho\beta_{0}}(\frac{1}{E_{\tau}}-\frac{1}{E_{\tau}^{i}})\Bigg]
$$
\end{equation}
  
\begin{equation}
$$
E_{\tau}=\exp\Bigg[-\frac{\beta_{0}}{\beta_{1}}\bigg(1-e^{-\beta_{1}\rho Z}\bigg)+\ln\bigg(\frac{E_{\tau}^{i}}{E_{0}}\bigg)e^{-\beta_{1}\rho Z}\Bigg]E_{0}            
$$
\end{equation}
where;\\
\begin{equation}
$$
\left\lbrace
	\begin{aligned}
\beta &= \beta_{0}+\beta_{1}ln\bigg(\frac{E}{E_{0}}\bigg) \\
\beta_{0} &=1.2\times10^{-6}cm^{2}g^{-1} \\
\beta_{1} &=0.16\times10^{-6}cm^{2}g^{-1}
	\end{aligned}
	\right.
$$
\end{equation}
In these expressions, $Z$ is the distance traveled by the $\tau$ after its production,$E_{\tau}^{i}$ is the initial $\tau$ energy just at moment of its production with $E_{\tau}^{i}=0.8E_{\nu}$ and $E_{0}=10^{10}GeV$. 

\subsection{In the valley: The horizontal air shower created by tau lepton}
The four main decay channels of $\tau$ are [21]:
$$
\left\lbrace
	\begin{aligned}
\tau^{-} &\rightarrow e^{-}\bar{\nu}_{e}nu_{\tau} \: (17.84\%) \\
\tau^{-} &\rightarrow \mu^{-}\bar{\nu}_{\mu}nu_{\tau} \,(17.36\%) \\
\tau^{-} &\rightarrow 3h nu_{\tau}(+n\pi_{0}) \,(15.19\%) \\
\tau^{-} &\rightarrow h_{-} nu_{\tau}(+n\pi_{0}) \,(49.22\%)
	\end{aligned}
	\right.
$$

\section{CONEX code}
CONEX [22] is a hybrid simulation code, implemented as an option in the CORSIKA \cite{Réf 1} code, which is considered one of the best EAS simulation codes using the Monte Carlo method. CONEX combines Monte Carlo simulation (MC) with a fast numerical solution of cascade equations (CE) for the resulting distributions of high energy secondary particles. It enables fast simulations of air showers, including fluctuations. \\
For a primary particle, a given energy and zenith angle, the energy deposition profile as well as the longitudinal profiles of charged particles (electrons, positons, muons) and gammas photons are calculated.\\ 
The parameters of the air shower simulation, profiles and adjustment results are written to a ROOT file.
\section{Results and discussions}
\subsection{Tau lepton production}

\begin{figure}[h]
	\begin{center}
		\includegraphics[scale=0.55]{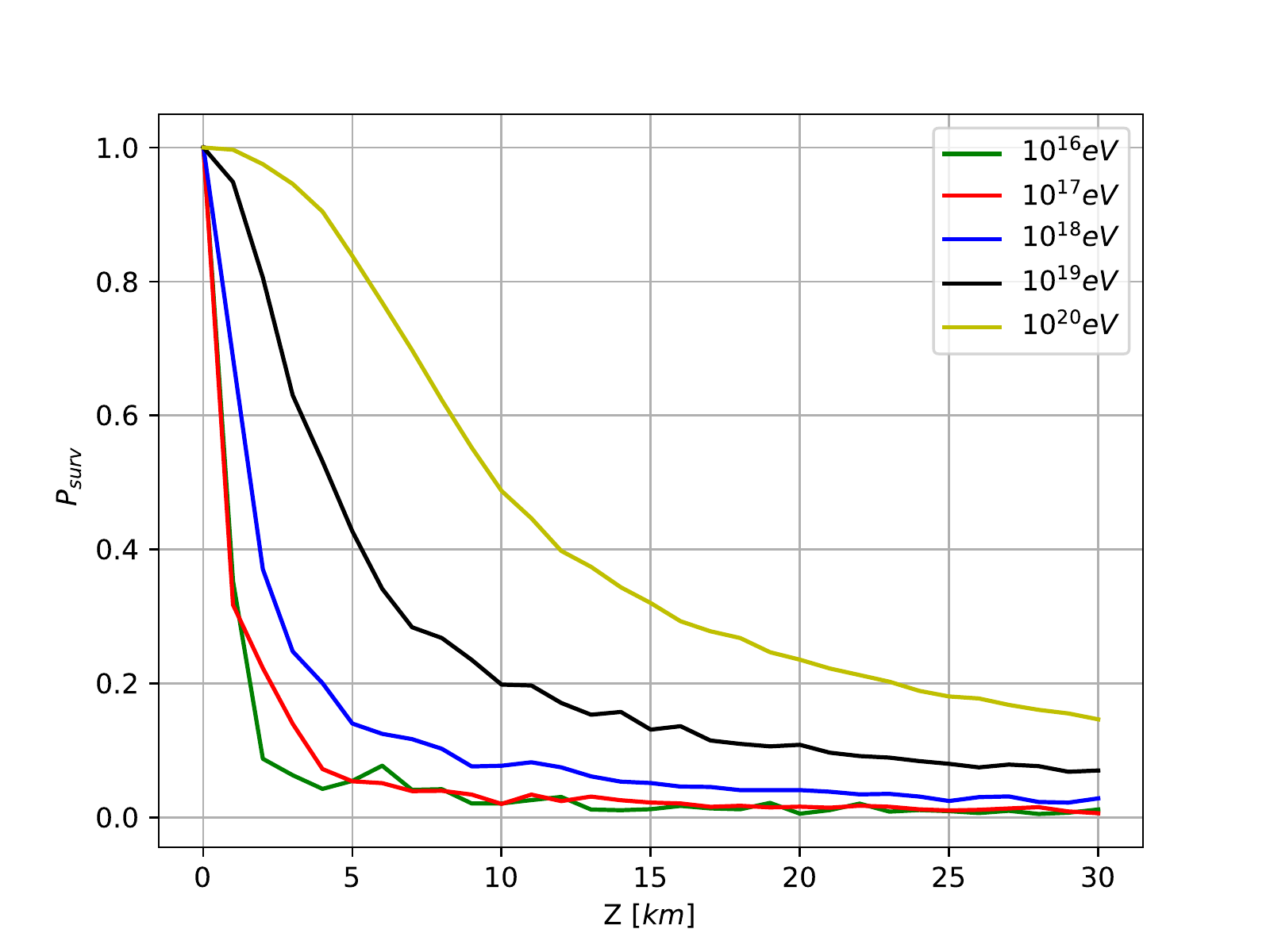}
	\end{center}
	\caption{Tau lepton survival probability as a function of the distance $Z$ travelled $\tau$ after its production, for five energies  $\nu_{\tau}$ incident}
\end{figure}

The $\tau$ lepton survival probability in relation to the distance traveled by the $\tau$ after its production in the mountain, for primary $\nu_{\tau}$ energies ranging from $10^{16} eV$ to $10^{20} eV$ is shown in Figure 1. We clearly notice the probability of survival for $\tau$ increases with the energy of $\nu_{\tau}$. 

In figure 2, we plot the energy distribution of the tau lepton produced from mono-energetic neutrino of $10^{20} eV$ as a function of the distance travelled by $\tau$ for a total rock depth equal to $25 km$. The tau is created at a depth of $D_{int}=21.5km$, and its enery decreases from $ E_{\tau}^{i}=8\times10^{19} eV$ to $E_{\tau}^{i}=2\times10^{19} eV$ just at moment of its emergence.

\begin{figure}[h]
	\begin{center}
		\includegraphics[scale=0.55]{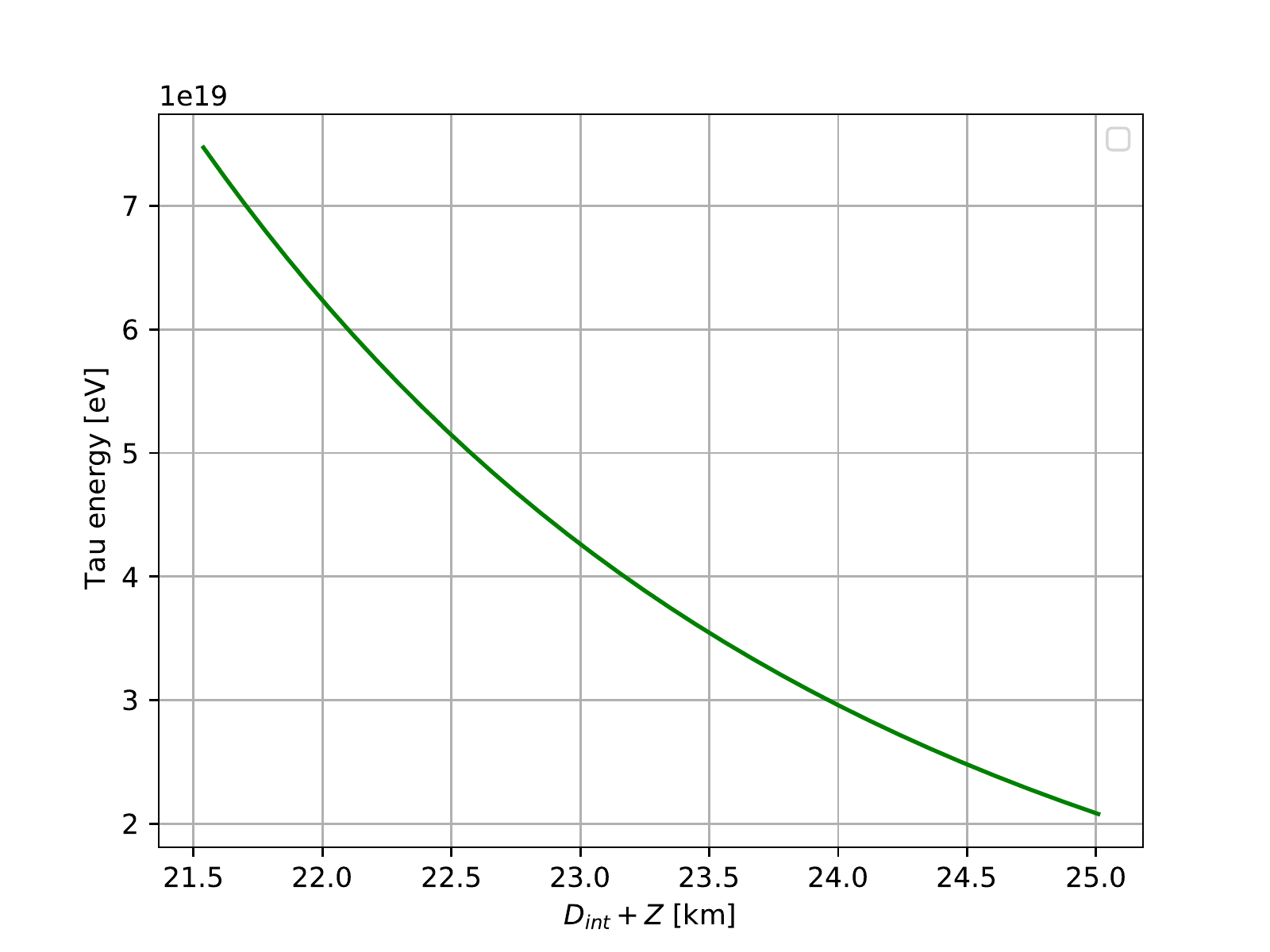}
	\end{center}
	\caption{The tau lepton energy as function of the distance travelled in the mountain, with:$E_{\nu}=10^{20}eV$ and $D_{int}=21.5km$}  
\end{figure}

\subsection{Decay of Tau lepton}
Figure 3 illustrates the results obtained in the case of a neutrinos energy of $10^{19} eV$. For different valley widths, the number of $\tau$ emerging from the rock and the number of $\tau$ decaying at the level of the valley.\\
It is observed that there are two regions. The first region exhibits a decrease in the number of $\tau$ for small mountain thicknesses. This is due to the fact that neutrino conversion requires enough material. Then, with the increase in the width of the mountain, a plateau appears. \\
This plateau is the result of two antagonistic effects: a greater probability of production of $\tau$ linked to the increase in matter combined with the production of $\tau$ farther and farther from the opposite end of the mountain, that is, with a lower probability of escape.\\
The minimum required width can be deduced from this figure: approximately $5 km$ for a neutrinos energy of $10^{19} eV$. It is also observed that, for $10^{19} eV$, a valley length of $10 km$ allows the decay of all born $\tau$ (the black and red curves merge).\\
Figures 4 and 5 illustrate the results obtained in the case of neutrino energies of $10^{18} eV$ and $10^{20} eV$. \\
The same effects as above are observed, with, in the case of $10^{20} eV$ as an example, an expansion of the required valley and mountain widths. The opposite effect is observed for $10^{18} eV$ (A contraction in the required distance) with, in this case, an increase in statistical fluctuation.

\begin{figure}[h!]
	\begin{minipage}[b]{0.40\linewidth}
		\centering \includegraphics[scale=0.42]{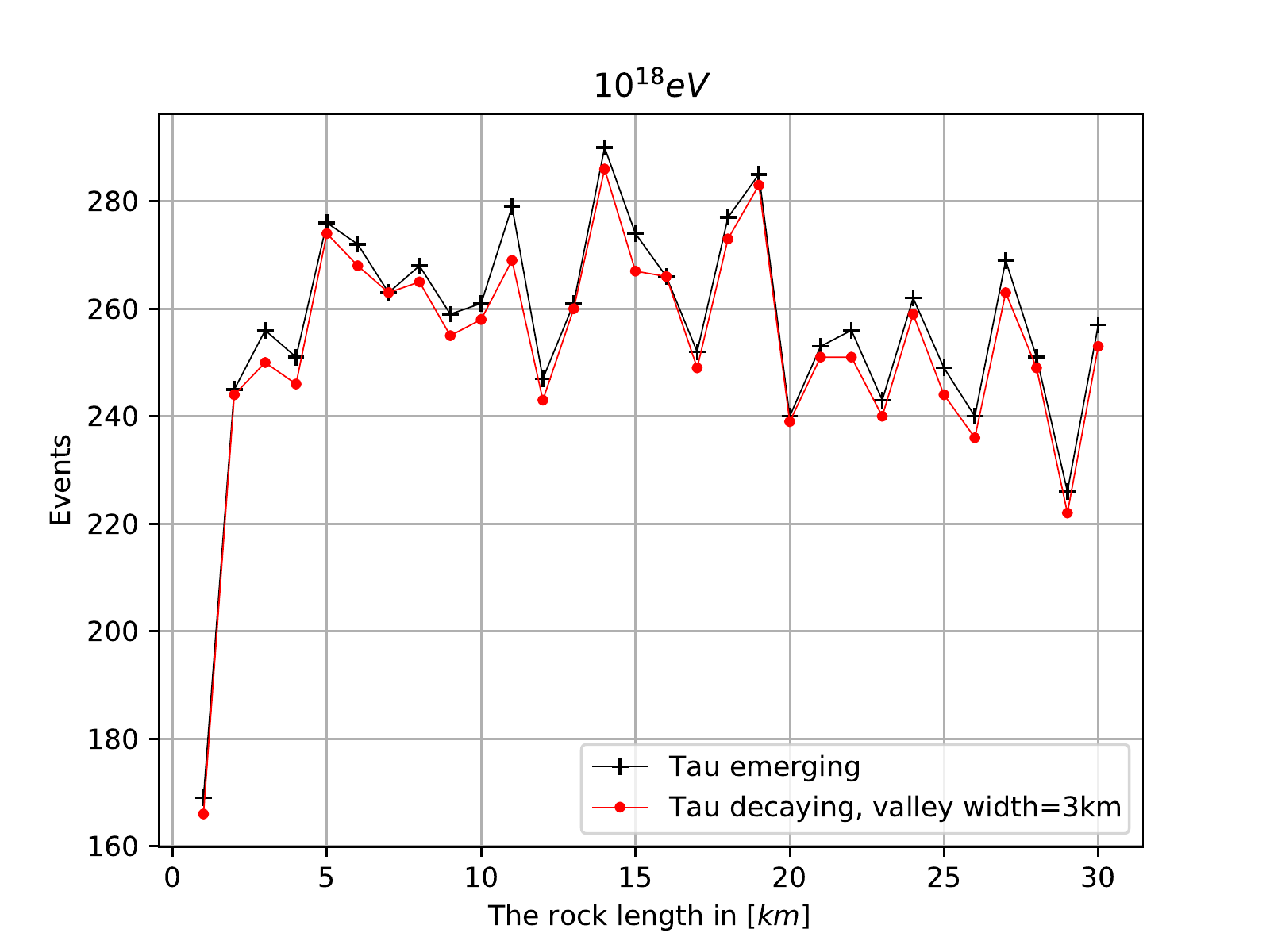}
		%\caption{\it Légende 1}
	\end{minipage}\hfill
	\begin{minipage}[b]{0.48\linewidth}	
		\centering \includegraphics[scale=0.42]{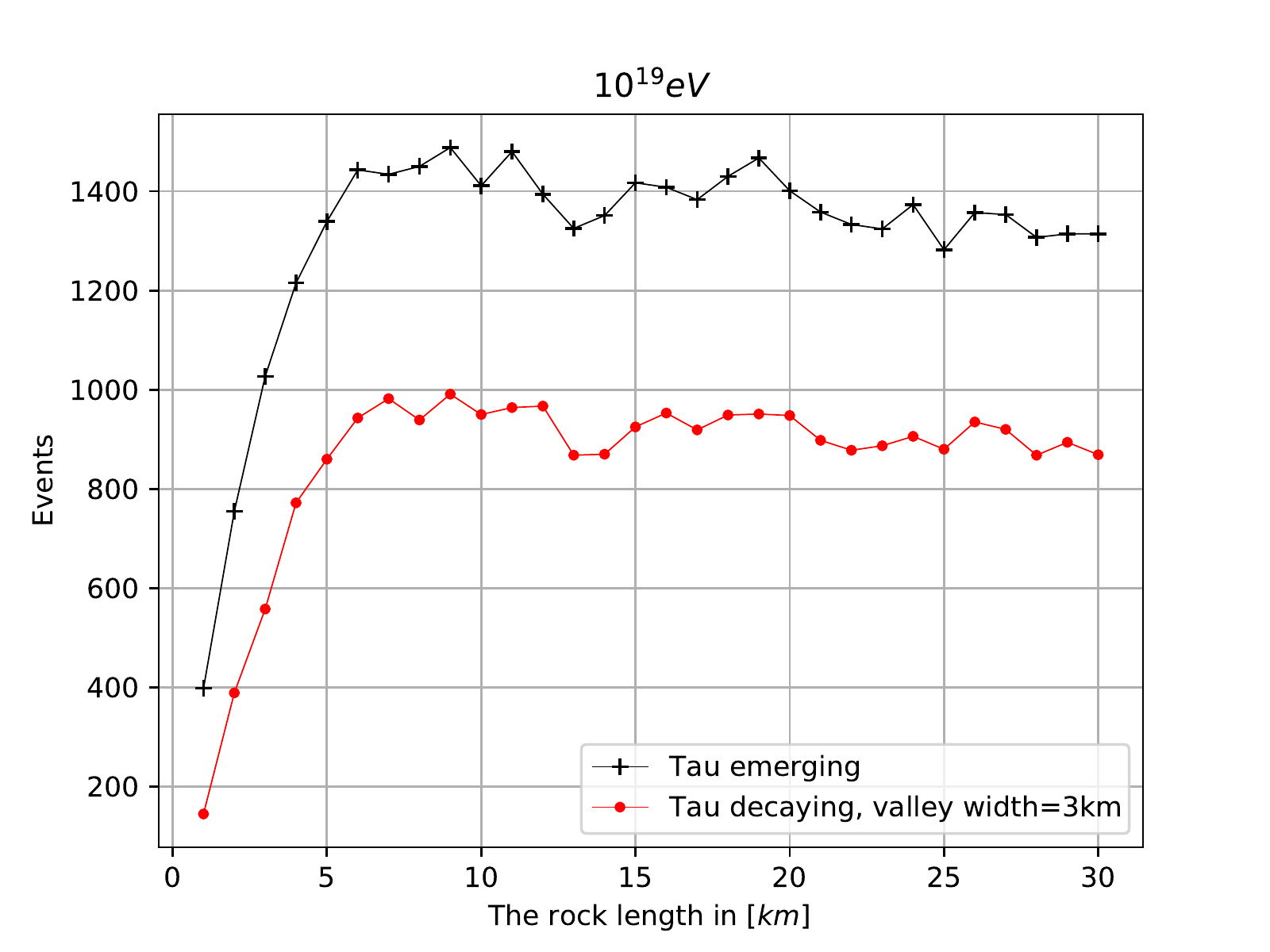}
		%\caption{Légende 2}
	\end{minipage}
	\begin{minipage}[b]{0.40\linewidth}
		\centering \includegraphics[scale=0.42]{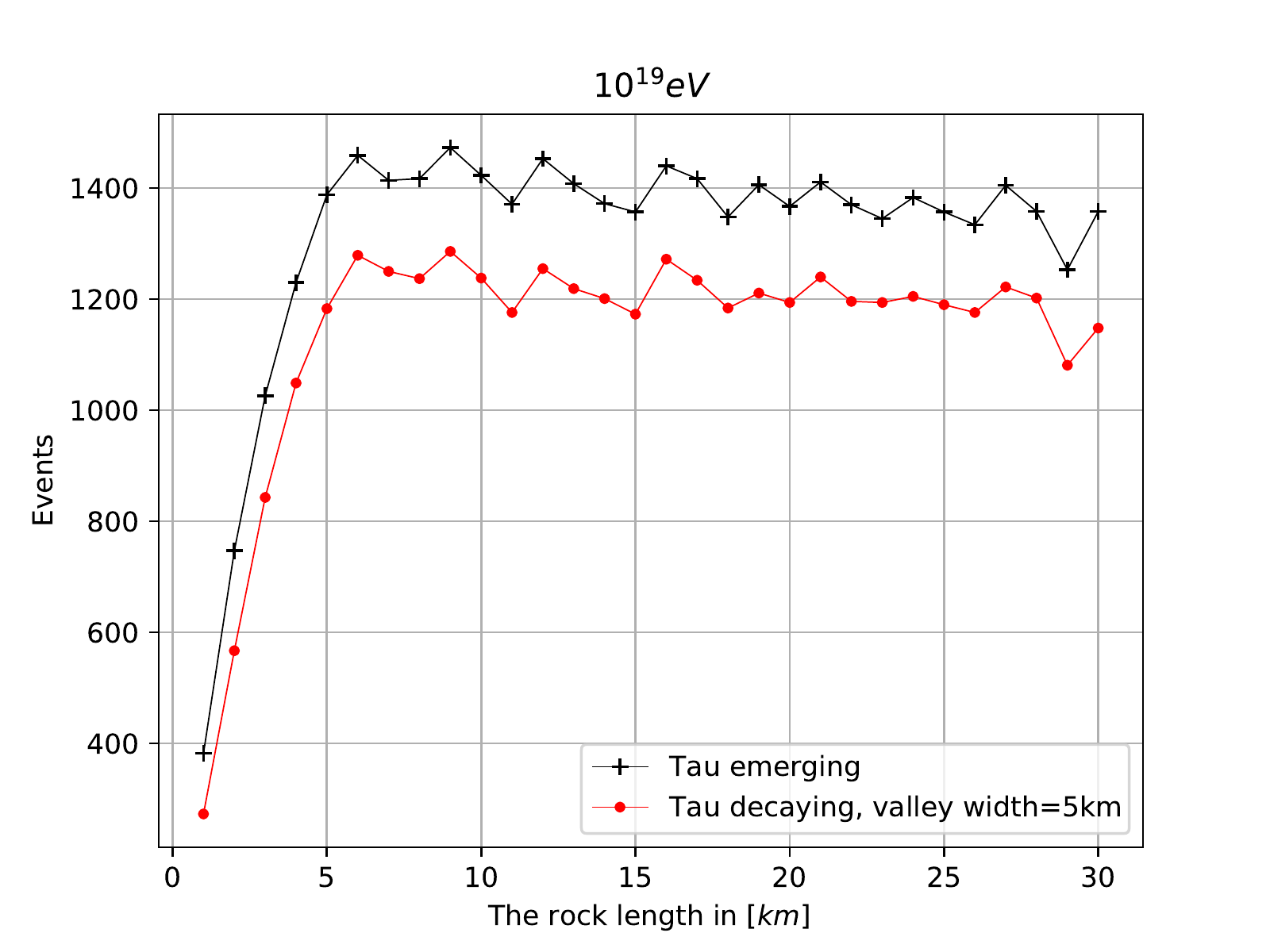}
		%\caption{Légende 3}
	\end{minipage}\hfill
	\begin{minipage}[b]{0.48\linewidth}	
		\centering \includegraphics[scale=0.42]{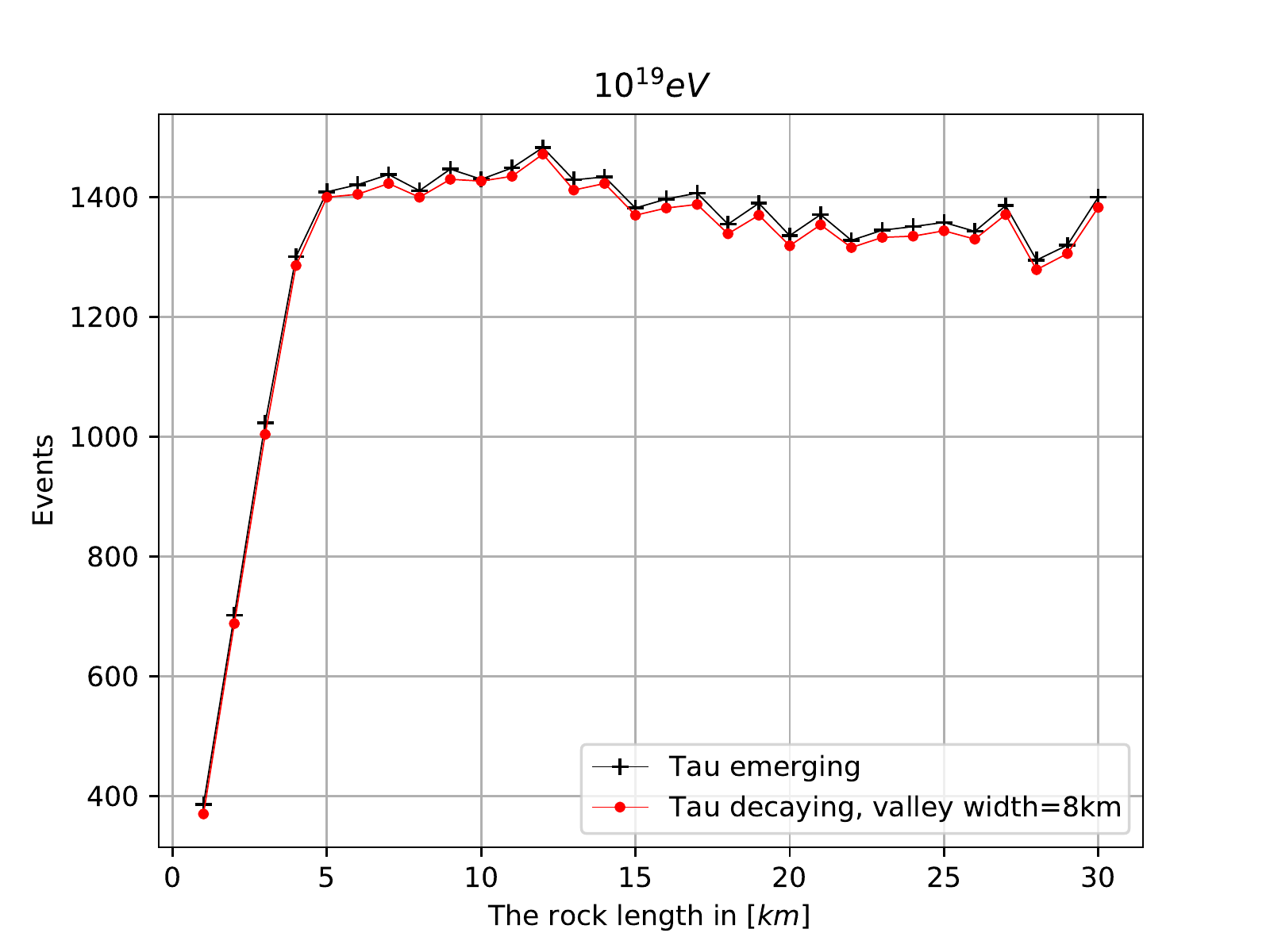}
		%\caption{Légende 4}
	\end{minipage}
	\caption{Number of tau leptons (emerging with black color and decaying with red color ) as function of the rock length for two values of tau neutrino energies $10^{18} eV $ and $10^{19} eV $ for different values of the valley widths (3, 5 and 8 kms)}
\end{figure}

\begin{figure}[h!]
	\begin{minipage}[b]{0.40\linewidth}
		\centering \includegraphics[scale=0.42]{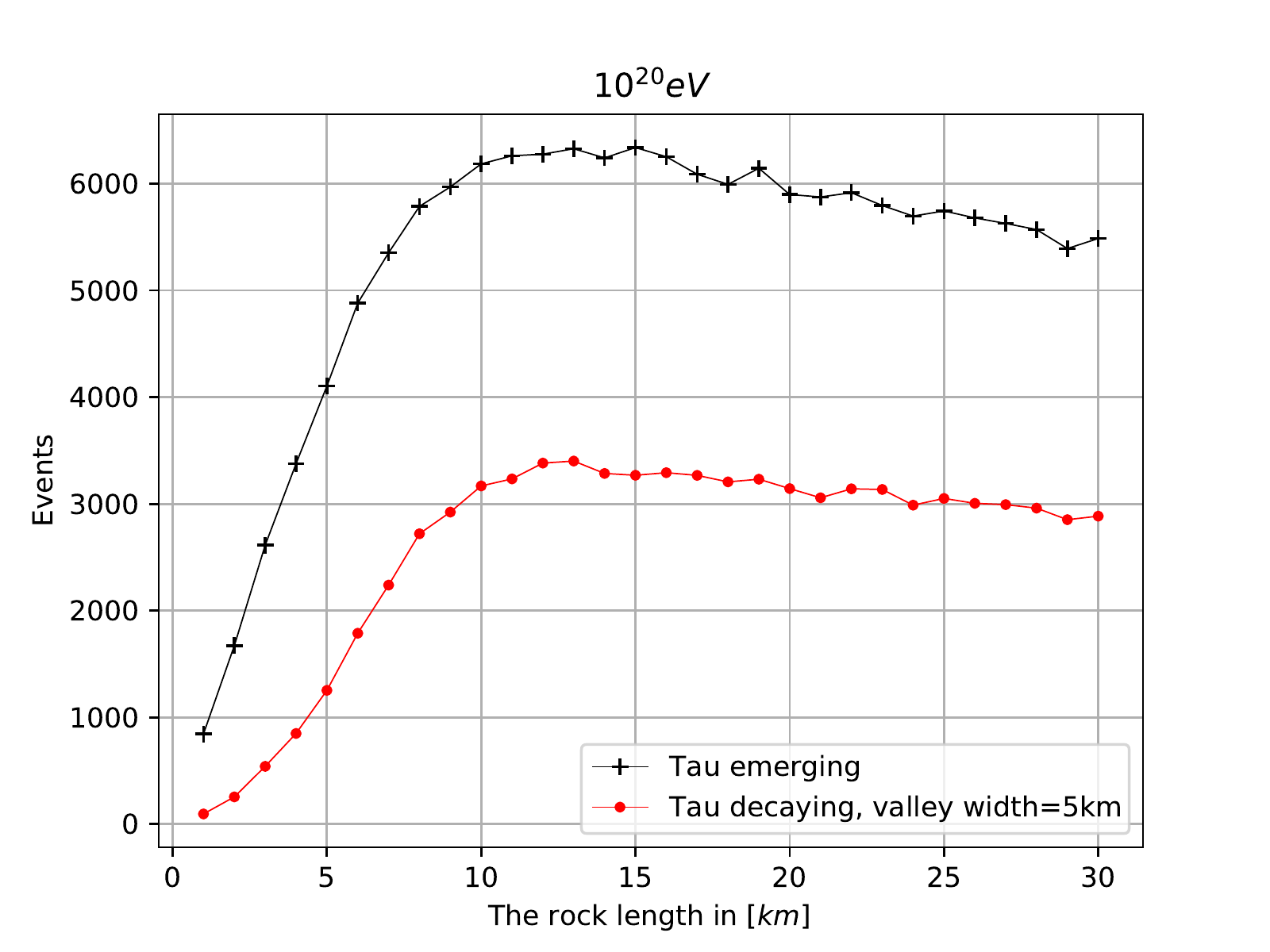}
		%\caption{\it Légende 1}
	\end{minipage}\hfill
	\begin{minipage}[b]{0.48\linewidth}	
		\centering \includegraphics[scale=0.42]{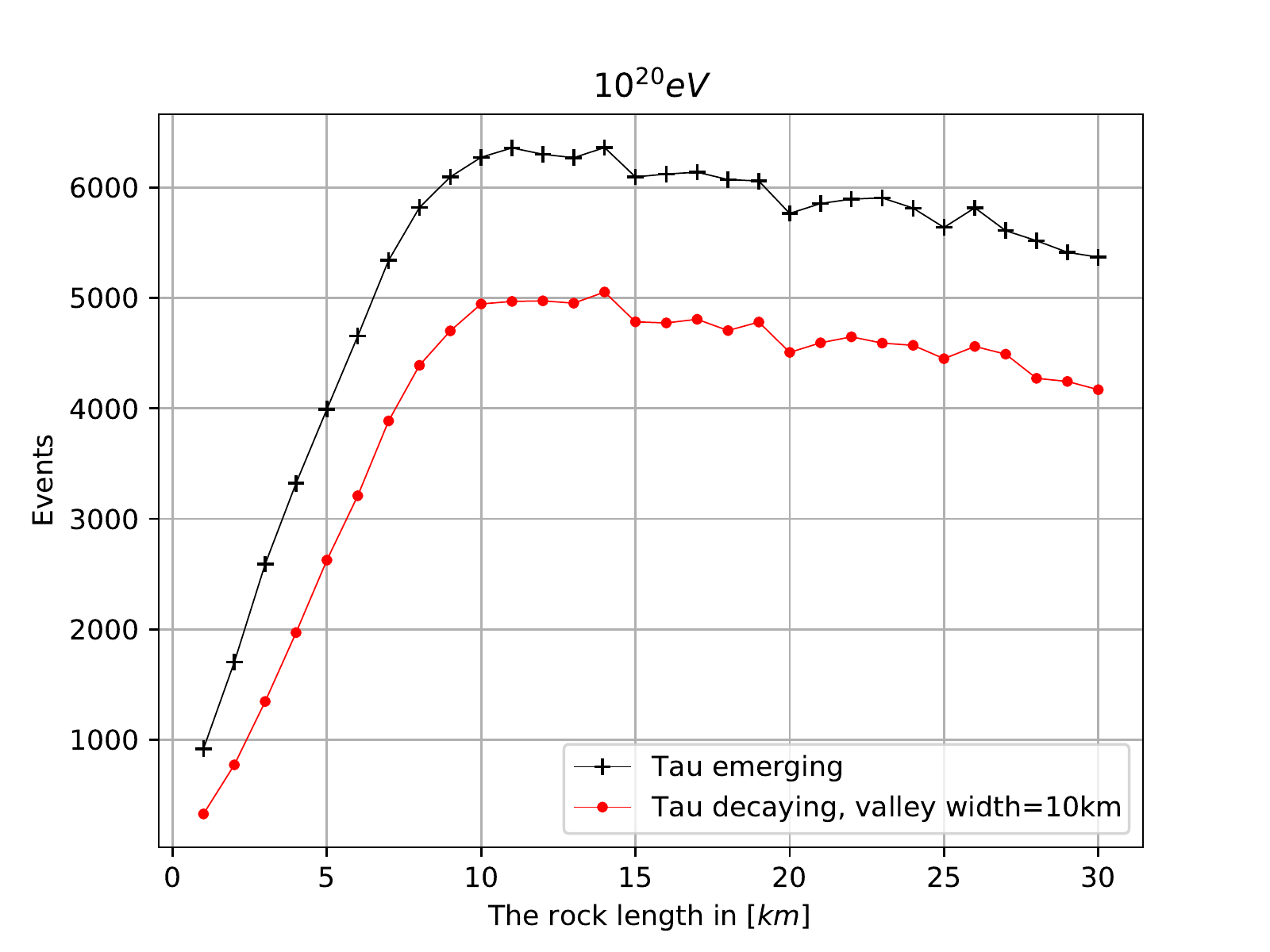}
		%\caption{Légende 2}
	\end{minipage}
	\begin{minipage}[b]{0.40\linewidth}
		\centering \includegraphics[scale=0.42]{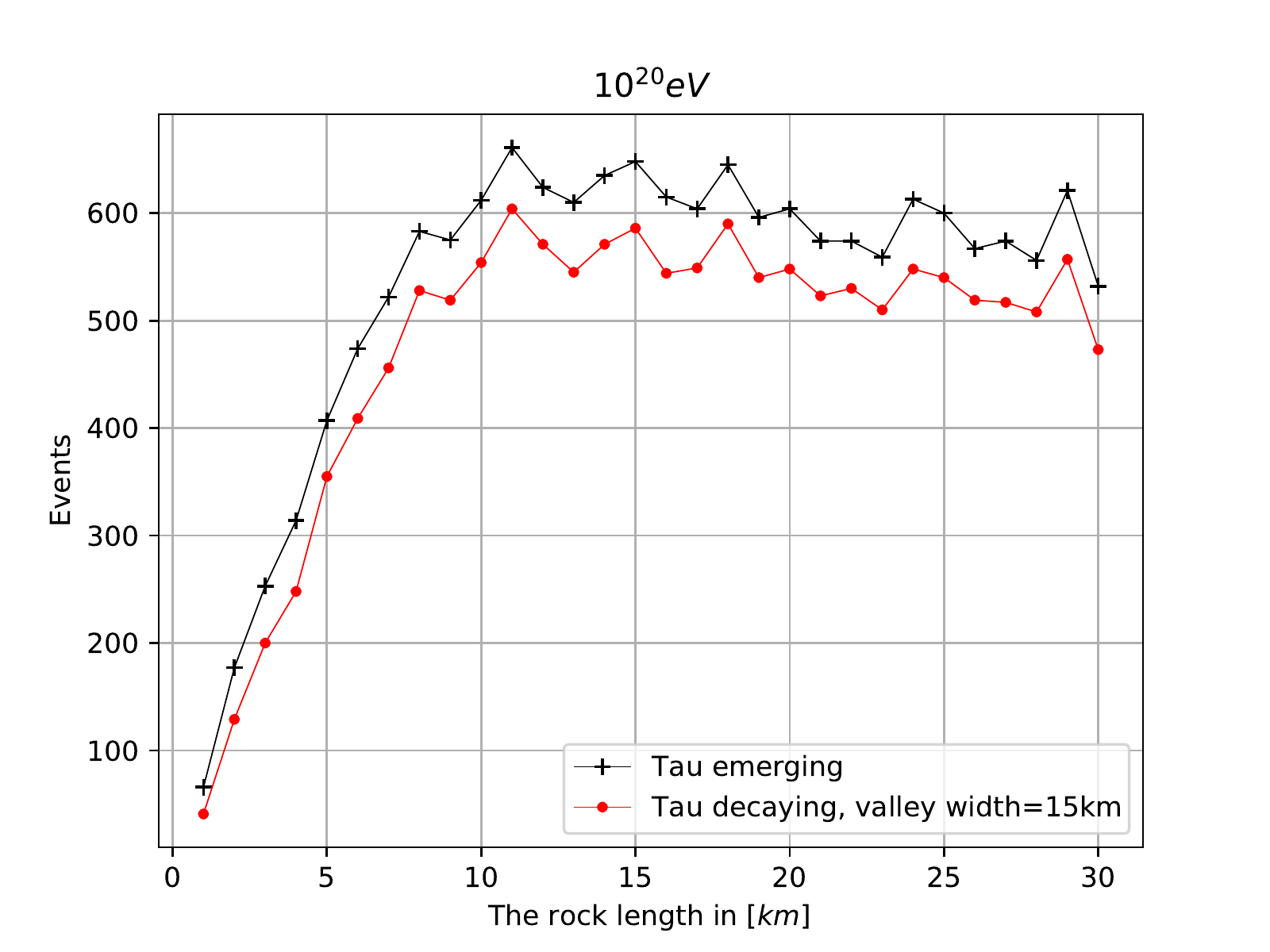}
		%\caption{Légende 3}
	\end{minipage}\hfill
	\begin{minipage}[b]{0.48\linewidth}	
		\centering \includegraphics[scale=0.42]{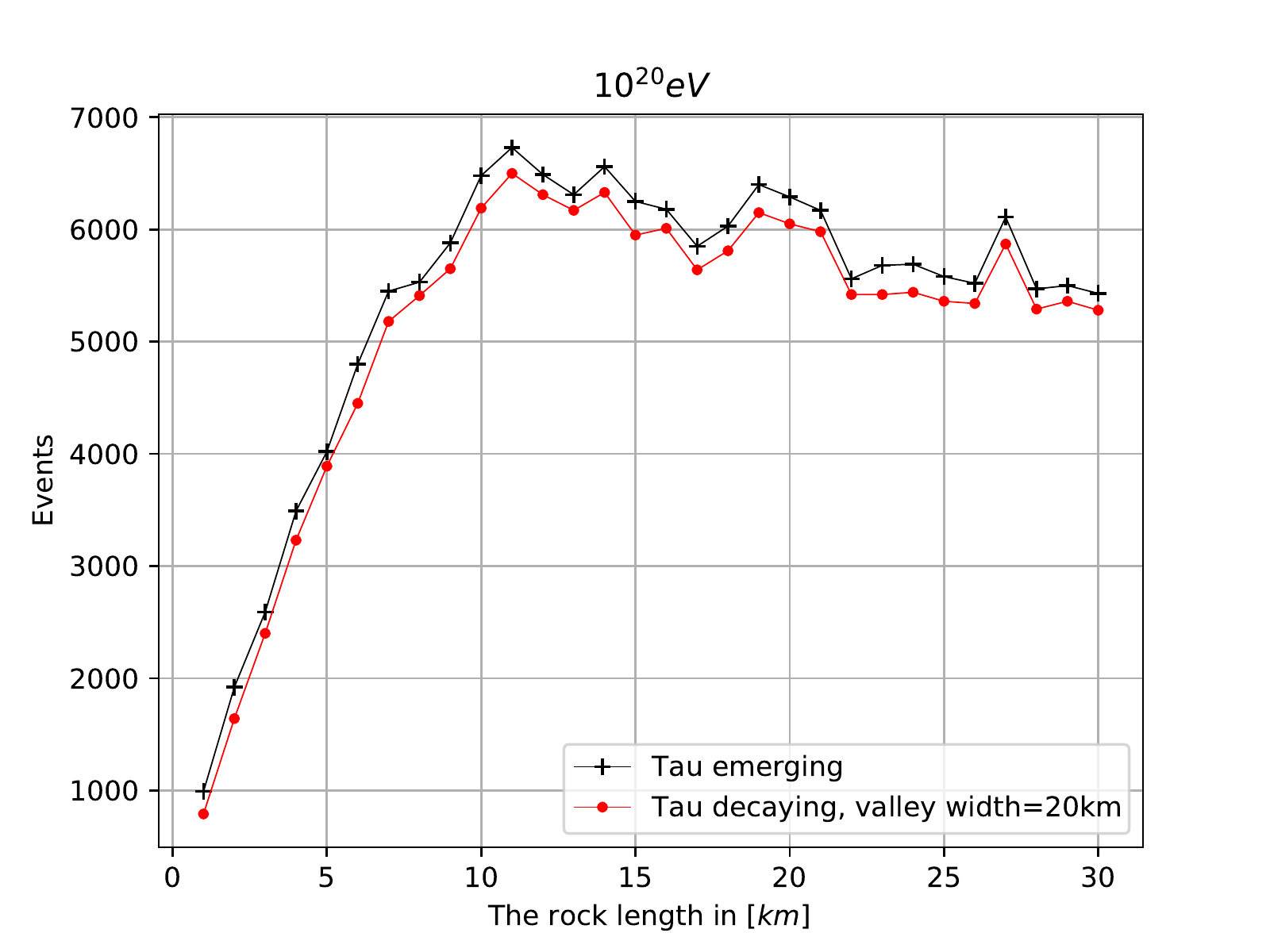}
		%\caption{Légende 4}
	\end{minipage}
	\caption{Number of tau leptons (emerging with black color and decaying with red color ) as function of the rock length for the tau neutrino energy to $10^{20} eV $ for different values of the valley widths (5, 10, 15 and 20 kms)}
\end{figure}

\subsection{Extensive air shower of tau lepton in valley}
In order to simulate tau lepton air showers,CORSIKA was applied. By choosing the CONEX option, the parameters of the simulations were set as follows:\\
\\
Primary particle : $\tau$ \\
Energy of the primary particle : $7*10^{6} GeV$ to $10^{10} GeV$ \\
Zenith angle : $ \theta =80^{°}$ \\
Event number : 800 to 1000 \\
Hadronic interaction model : QGSJET01 \\
Electromagnetic Component : EGS4 \\
During each simulation, the Gaisser-Hillas parameters ($N_{max} $-number  of charged particles at the maximum of the shower, $X_{max}$ - maximum atmospheric depth of the shower) are recorded.\\

\begin{figure}[h!]
	\begin{minipage}[b]{0.40\linewidth}
		\centering \includegraphics[scale=0.42]{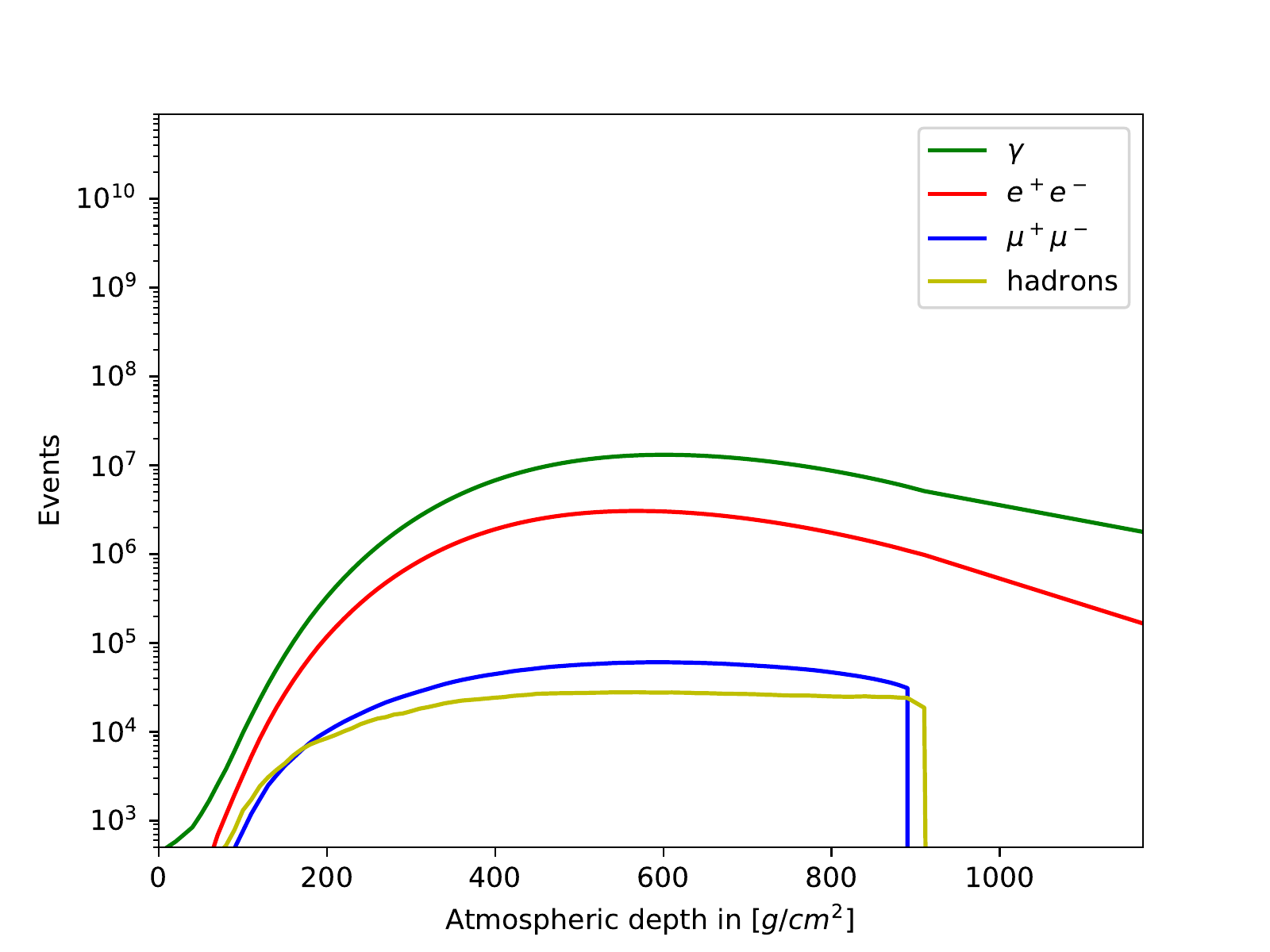}
		%\caption{\it Légende 1}
	\end{minipage}\hfill
	\begin{minipage}[b]{0.48\linewidth}	
		\centering \includegraphics[scale=0.42]{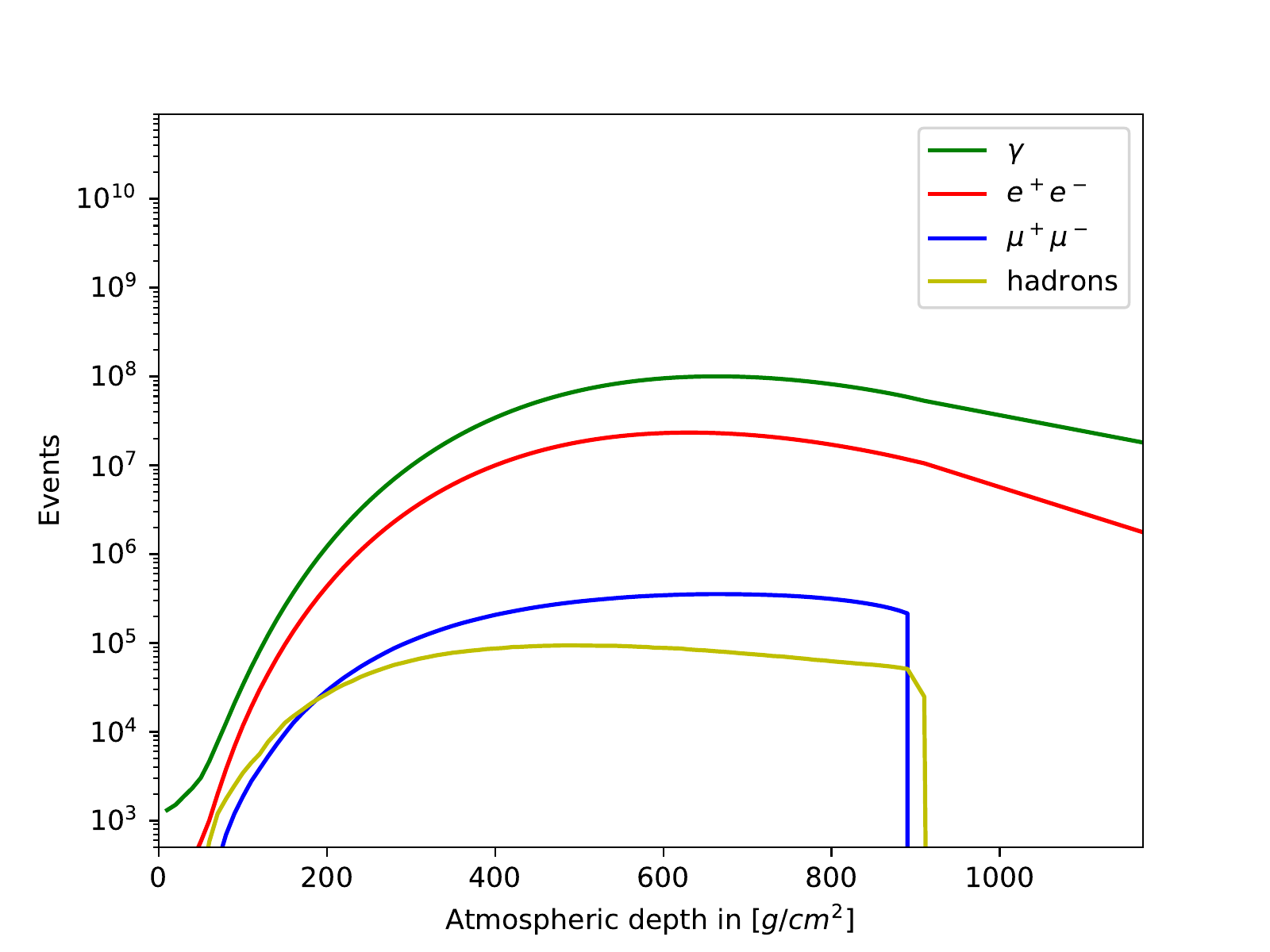}
		%\caption{Légende 2}
	\end{minipage}
	\begin{minipage}[b]{0.40\linewidth}
		\centering \includegraphics[scale=0.42]{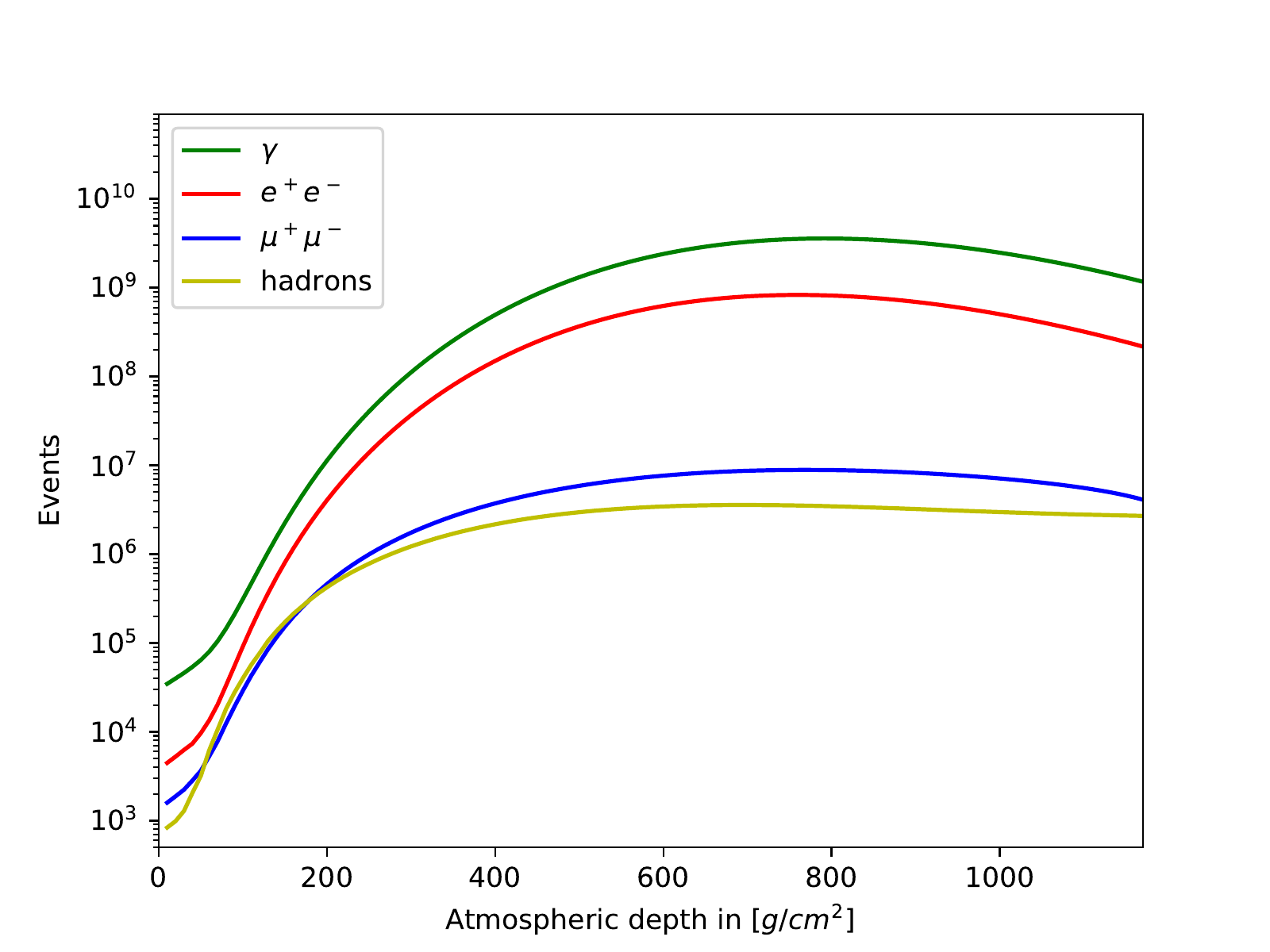}
		%\caption{Légende 3}
	\end{minipage}\hfill
	\begin{minipage}[b]{0.48\linewidth}	
		\centering \includegraphics[scale=0.42]{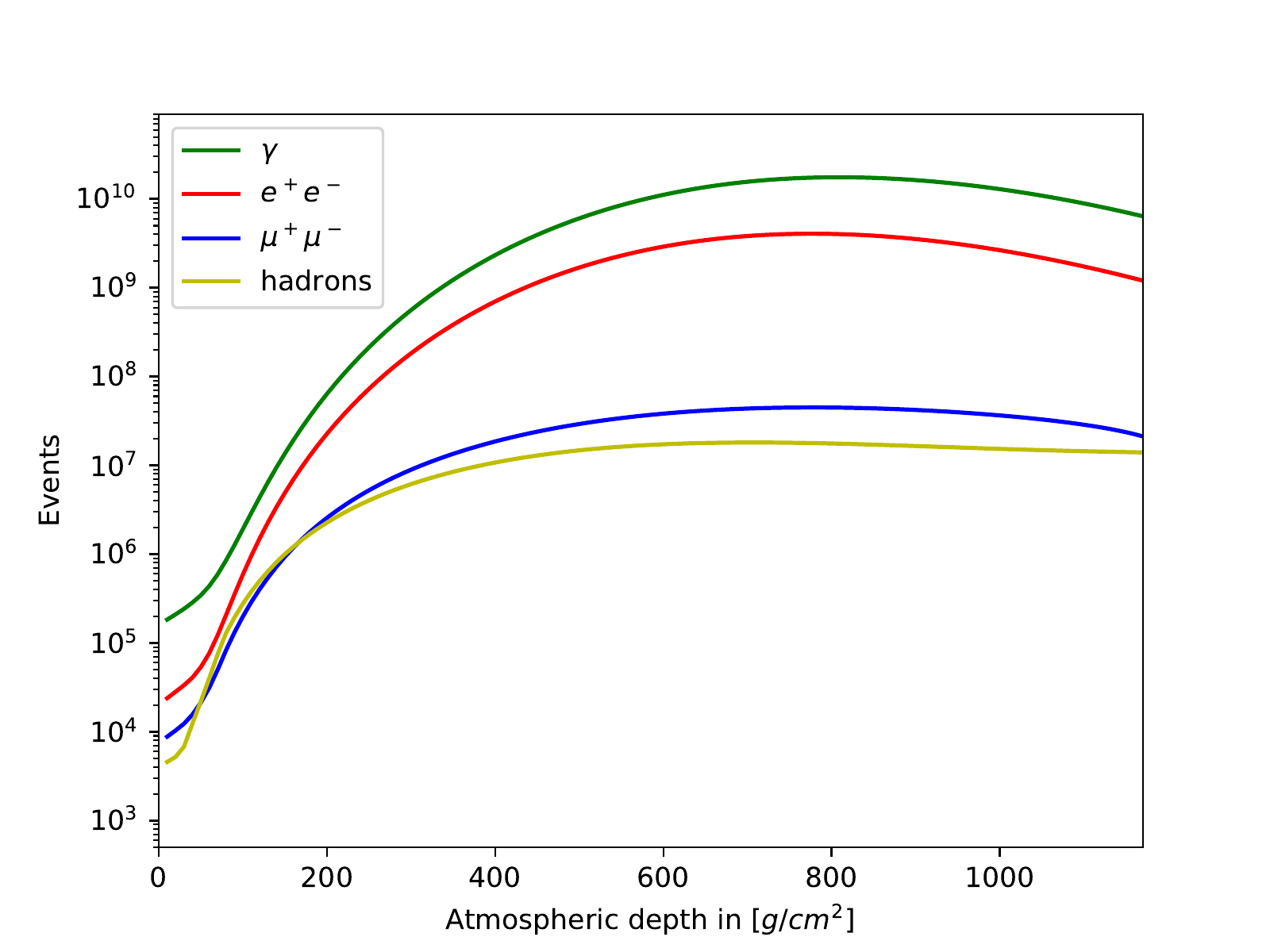}
		%\caption{Légende 4}
	\end{minipage}
	\caption{The longitudinal profile for differents particles daughters from showers initiated by tau leptons as function of atmospheric depth.
	Top panels( left:$E_{\tau}=7\times10^{15} eV$, right:$E_{\tau}=5.5\times10^{16} eV$), Bottom panels ( left:$E_{\tau}=2\times10^{18} eV$, right:$E_{\tau}=10^{19} eV$) }
\end{figure}

In figure 5 we plot as function of atmospheric depth (X in $g/cm^{2}$) the longitudinal particles distribution for differents particles daughters from showers initiated by tau leptons. For all energies, we notice that the $N_{\gamma}$ is the highest followed by $N_{e\pm}$. The $N_{\mu\pm}$ and $N_{hadrons}$ represents $\sim0.5\%$ of the total number of particles in the shower.\\ 
As the tau energy increases, the peak $N(X)$ becomes higher and shifts to large values of depth atmospheric. The values of the peak $N_{max}$ corresponds to a value of $X=X_{max}$.\\
Figure 6 shows the variation of the depth atmospheric shower maximum $X_{max}$ with the tau enery initiated showers. In the enery ranges from $10^{15} eV$ to $10^{19} eV$, we notice that the relationship between  $X_{max}$ and $E_{\tau}$ is quasilinear.

\begin{figure}[h!]
	\begin{center}
		\includegraphics[scale=0.55]{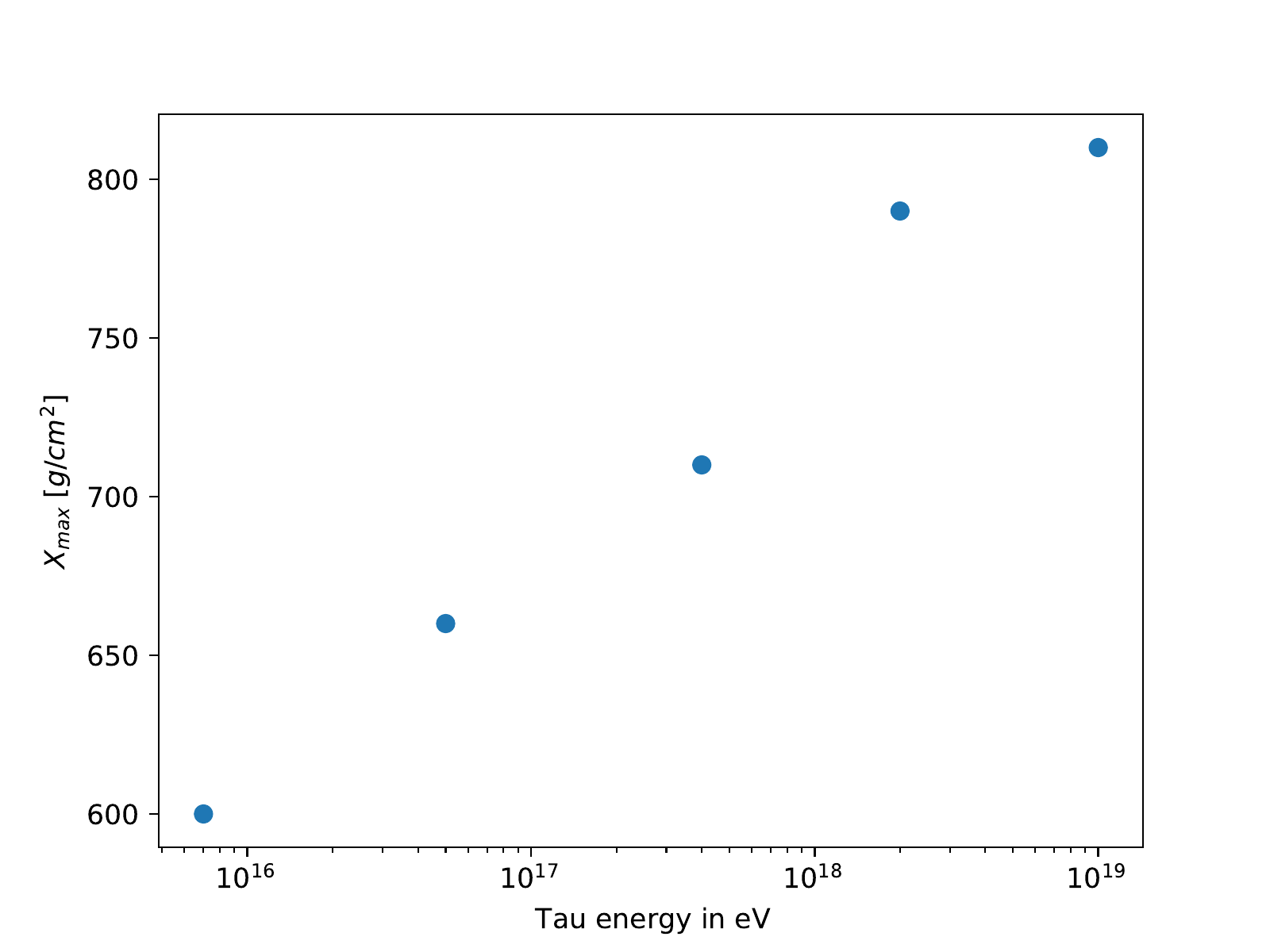}
	\end{center}
	\caption{The $X_{max}$ as a function of energy showers initiated by tau leptons }
\end{figure}

\begin{figure}[h!]
	\begin{center}
		\includegraphics[scale=0.55]{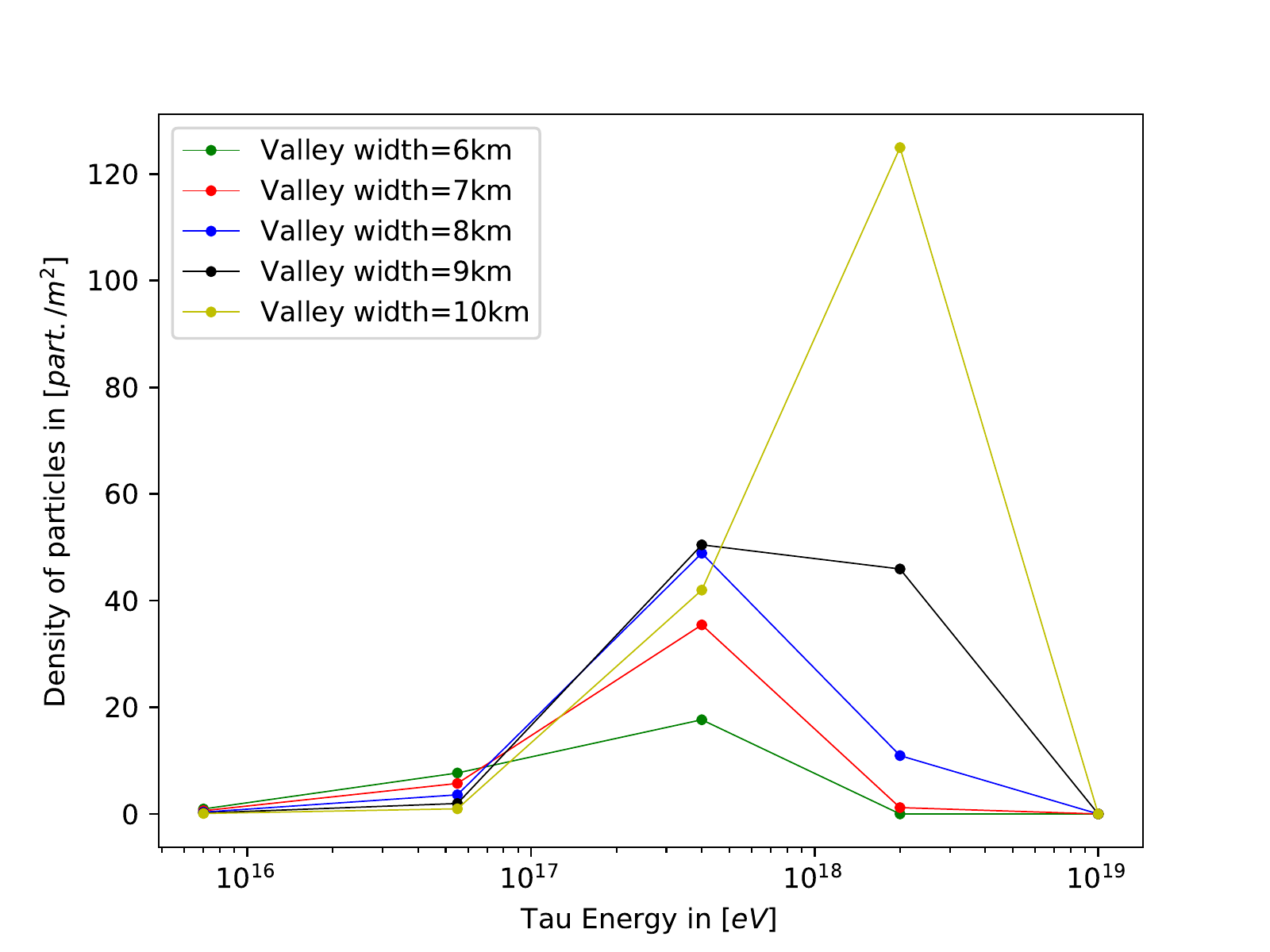}
	\end{center}
	\caption{Density in particules ($\gamma , e^{\pm}and \mu^{\pm}$) per $m^{2}$ as a function of incident tau energy for different detector positions (Valley width=6,7,8,9 and 10 kms)}
\end{figure}

The distribution of particles density in $Part./m^{2}$ for different tau lepton energy, for various detector positions (Valley witdh = 6, 7, 8, 9 and 10 kms) its shows in figure 7. 
For all values of valley width, the density of particles growth with energy, goes through a peak at $3\times10^{17}eV$, then it decreases, in except the valley width equal to 10km have a peak at $2\times10^{18}eV$. 
From the results we can deduce that the value of the valley width equal to 9 km allows the detection of the widest range of tau lepton energies from $10^{15}eV$ to $10^{19}eV$. 
Beyond a valley width greater than 10 km the density of the particles will take very small values for the energy range from 15 to $5\times10^{17}eV$ and the peak will shift towards an energy close to $10^{19}eV$ which prevents detection for lower energies.

\section*{Conclusions and perspectives}
	This research studied the use of a (mountain-valley) configuration as a possible outline for the detection of UHE tau neutrinos.
With Monte Carlo simulation, it was possible to determine the dimensions of the first mountain used to produce a maximum number of emerging tau leptons, which range from 5 km to 10 km.
The CONEX simulation enabled a detailed study of the air shower born from tau lepton, it allowed us to deduce the optimal position for the detection network 
which corresponds to the valley width equal to 9 kms. 
To allow the detection of these showers, the dimensions of the scintillators must be greater than $1 m^{2}$ because the density of the particles is of the order of a few particles per $m^{2}$ for tau energies lower than $10^{16} eV$.\\

%\end{multicols}

\begin{thebibliography}{15}
	\bibitem{Réf 1}
	D. Heck and al., Report FZKA 6019 (1998)
	\bibitem{Réf 2}
	J.A. Formaggio et G.P. Zeller, Rev. Mod. Phys., 48(2012)3
	\bibitem{Réf 3}
	D. Fargion,1,2 B. Mele1 and A. SALIS, The Astrophysical Journal, 517 : 		725È733(1999)
	\bibitem{Réf 4}
	A. Letessier-Selvon, ASP Conference Series, Vol. 241(2001)
	\bibitem{Réf 5}
	X. Bertou, P. Billoir, O. Deligny, C. Lachaud, A. Letessier-Selvon. Astroparticle Physics, Elsevier,2002, 17, pp.183-193
	\bibitem{Réf 6}
	L. Fengab, P. Fisher, F. Wilczek and al., Phys. Rev. Lett. 88,161102 (2002)
	\bibitem{Réf 7} 
	K. Kotera,PoS(ICRC2021), vol. 395, p. 1181, (2021)
	\bibitem{Réf 8} 
	A. Zeolla and al.,PoS(ICRC2021), vol. 395, p. 1072, (2021)
	\bibitem{Réf 9} 
	A. Brown,PoS(ICRC2021), vol. 395, p. 1179, (2021)
	\bibitem{Réf 10} 
	ARA Collaboration, P. Allison, and et al., , Astroparticle Physics 35 457–477 (2012).
	\bibitem{Réf 11}  
	S. W. Barwick and et al., (2014), , arXiv:1410.7369.
	\bibitem{Réf 12}  A. Olinto et al., (2020), arXiv:2012.07945 [astro-ph.IM].
	\bibitem{Réf 13}  L. Wiencke and al., PoS
ICRC2019, 466 (2020), arXiv:1909.12835 [astro-ph.IM].
	\bibitem{Réf 14}
	B. Boussaha Détection de muons et de neutrinos-tau cosmiques USTHB (2013)
	\bibitem{Réf 15}
	F. Montanet, D. Lebrun, J. Chauvin, E. Lagorio, and P. Stassi Astrophys. Space Sci. Trans., 7, 369-372, (2011)
	\bibitem{Réf 16}
	D. Lebrun, F. Montanet, J. Chauvin, D-H. Koang, E. Lagorio and P. Stassi, Proceedings of The 31st ICRC, ŁÓDZ (2009)
	\bibitem{Réf 17}
	Kévin Payet. astro-ph.CO. Université Joseph-Fourier-Grenoble I, Français. ff tel-00451532v2f(2009)
	\bibitem{Réf 18}
	L. A. Anchordoqui, A. M. Cooper-Sarkar, D. Hooper and S. Sarkar, Phys. Rev. D 74 (2006)043008 [arXiv:hep-ph/0605086]
	\bibitem{Réf 19}
	O.B. Bigas et al., arxiv:astro-ph 0806.2126v2 (2008)
	\bibitem{Réf 20}
	Y. Huang et M.H. Reno, Phys. Rev. D72, (2005) 013005
	\bibitem{Réf 21}
	P.A. Zyla et al. (Particle Data Group), Prog. Theor. Exp. Phys. 2020,   
	083C01 (2020)
	\bibitem{Réf 22}
	T. Pierog, Hybrid Simulation and CONEX kernel, KIT (2014)	
	\end{thebibliography}
\end{document}